\documentclass[a4paper,12pt]{article}
\usepackage{amssymb}
\usepackage{epsf,amsmath}
\usepackage[dvips,usenames]{color}
\usepackage{graphicx}
\usepackage{color}
\usepackage{colortbl}

\newlength{\dinwidth}
\newlength{\dinmargin}
\newcommand{\spur}[1]{\not\! #1 \,}

\begin{document}
\title{\bf Rare decay $\pi^{0}\to e^+ e^-$ constraints on the  light CP-odd Higgs in NMSSM }
\author{Qin Chang$^{a,b}$, Ya-Dong Yang$^{a,c}$\\
{$^{a}$\small  Institute of Particle Physics, Huazhong Normal University, Wuhan,
Hubei  430079, P. R. China}\\
{ $^b$\small Department of Physics, Henan Normal University,
Xinxiang, Henan 453007, P.~R. China}\\
{$^{c}$\small Key Laboratory of Quark $\&$ Lepton Physics, Ministry of Education, P.R. China }
\\} \maketitle
\bigskip\bigskip
\maketitle \vspace{-1.5cm}
\begin{abstract}
 We constrain the light CP-odd Higgs  $A_{1}^{0}$ in  NMSSM via the  rare decay $\pi^{0}\to e^{+}e^{-}$. It is shown that the possible $3\sigma$ discrepancy between theoretical predictions and the recent KTeV measurement of ${\cal B}({\pi}^{0}\to e^+ e^-)$ cannot be  resolved when the  constraints from  $\Upsilon\to\gamma A_1^0$,   $a_{\mu}$ and $\pi^{0}\to \gamma \gamma$ are combined.  Furthermore, the combined constraints also exclude the  scenario involving  $m_{A_1^0}=214.3$ MeV, which is  invoked to explain the anomaly in the $\Sigma^{+}\to p \mu^{+}\mu^{-}$ decay found by the HyperCP Collaboration.

  \end{abstract}

\noindent {\bf PACS Numbers: 13.25.Cq, 14.80.Cp}

\newpage
\section{Introduction}
Theoretically, the rare decay $\pi^{0}\to e^+ e^-$ starts at the one loop
level in the standard model (SM), which has been extensively studied
\cite{Drell,  Bergstrom1, Babu,Bergstrom2, Ametller, Savage, Gomez, Knecht, Dorokhov,P2ll}
since the first investigation in QED  by Drell~\cite{Drell}. It is
nontrivial to make precise predictions of the branching ratio
 ${\cal B}_{SM}(\pi^{0}\to e^+ e^-)$ because its sub-process involves the
$\pi^{0}\to \gamma^{*}\gamma^{*}$ transition form factor.
 In Refs.\cite{ Bergstrom1, Babu,Bergstrom2,Ametller}, the decay was
 studied via the Vector-Meson Dominance (VMD) approach, where the results are in good agreement
 with each other and converge  in ${\cal B}(\pi^{0}\to e^+ e^-)\sim 6.2-6.4\times 10^{-8}$.
 By using the measured  value of
 ${\cal B}(\eta\to\mu^+\mu^-)$ to fix the  counterterms of the chiral amplitude in Chiral Perturbation Theory (ChPT),
Savage {\it et al}.  predicted
${\cal B}(\pi^{0}\to e^+ e^-)= (7\pm1)\times 10^{-8}$ \cite{Savage}.
Using a procedure similar to that used in Ref.\cite{Savage} (although
with an updated measurement of ${\cal B}(\eta\to\mu^{+}\mu^{-})$),
 Dumm and Pich predicted
$(8.3\pm0.4)\times10^{-8}$~\cite{Gomez}. Alternatively,
using  the lowest meson dominance (LMD) approximation to the large-$N_{c}$
spectrum of vector meson resonances to fix  the counterterms,  Knecht {\it et al}. predicted
$(6.2\pm0.3)\times10^{-8}$~\cite{Knecht}, which is about $4\sigma$ lower than the value predicted by
Ref.\cite{Gomez}  but which agrees with the others.
Most recently, using a dispersive approach to the amplitude and the
experimental results of the CELLO~\cite{cello} and CLEO~\cite{cleo1}
Collaborations for the pion transition form factor, Dorokhov and
Ivanov~\cite{Dorokhov} have found that
\begin{equation}\label{SMresult}
{\cal B}_{SM}(\pi^{0}\to e^+
e^-)\,=\,(6.23\,\pm\,0.09)\times10^{-8},
\end{equation}
which is consistent with most  theoretical predictions of
${\cal B}_{SM}(\pi^{0}\to e^+ e^-)$ in the literature.
Moreover, their prediction that ${\cal B}(\eta\to \mu^+
\mu^-)=(5.11\pm0.2)\times 10^{-6}$ agrees with the experimental data (which gives a value of
$(5.8\pm0.8)\times 10^{-6}$~\cite{PDG}).

Experimentally, the accuracy of the measurements of the decay has
increased significantly since the first $\pi^{0}\to e^{+}e^{-}$ evidence was observed by the
Geneva-Saclay group \cite{Fischer} in 1978 with
 ${\cal B}_{SM}(\pi^{0}\to e^+ e^-)=(22^{+24}_{-11})\times 10^{-8}$.
  A detailed summary of the experimental situation can be found
 in Ref.\cite{Rune}.  Recently, using the
complete data set from KTeV E799-II at Fermilab, the KTeV Collaboration
has made a precise measurement of the $\pi^{0}\to e^+ e^-$
branching ratio~\cite{KTeV}
\begin{equation}\label{Expresult}
{\cal B}^{no-rad}_{KTeV}(\pi^{0}\to e^+
e^-)\,=\,(7.48\,\pm\,0.29\,\pm\,0.25)\times10^{-8},
\end{equation}
after  extrapolating the full radiative tail beyond
$(m_{e^+e^-}/m_{\pi^0})^2>0.95$ and scaling their result back up by
the overall radiative correction of $3.4\%$.

As was already noted in Ref.~\cite{Dorokhov}, the SM
prediction given in Eq.(\ref{SMresult}) is $3.3\sigma$  lower
than the  KTeV data.  The authors have also compared
their result with estimations made by various approaches  in the literature and
found good agreements. Further analyses have found  that QED
radiative contributions~\cite{Dorokhov2} and mass
corrections~\cite{Dorokhov1} are at the level of a few percent and are therefore
unable to reduce the discrepancy.   Although the discrepancy might be due to
 hadronic dynamics that are as of yet unknown,  it is equally possible that
 this discrepancy is caused by the effects of new physics (NP). In this Letter
 we will study the latter possibility.

As is known that leptonic decays of pseudoscalar mesons are
sensitive to pseudoscalar weak interactions beyond the SM. Precise
measurements and calculations of these decays will offer sensitive
probes for NP effects at the low energy scale. Of particular
interest to us is the rare decay $\pi^{0}\to e^{+} e^{-}$, which
could proceed at tree level via a flavor-conserving process induced
by a light pseudoscalar Higgs boson $A^{0}_{1}$ in the
next-to-minimal supersymmetric standard model (NMSSM)~\cite{NMSSM}.
We will  look for a region of the parameter space of NMSSM that
could resolve the aforementioned discrepancy of ${\cal B}(\pi^{0}\to
e^{+}e^{-})$ at $1\sigma$. Then, we combine constraints from
$a_{\mu}$ and the recent searches for $\Upsilon(1S), (3S) \to \gamma
A^{0}_{1}$ by CLEO~\cite{cleo} and BaBar~\cite{babar}, respectively.

\section{The amplitude of $\pi^{0}\to e^{+} e^{-}$  in the SM and the NMSSM}

The NMSSM has generated considerable interest in the literature, which
extends the minimal supersymmetric SM (MSSM) by introducing  a
new Higgs singlet chiral superfield $\hat S$ to solve the known
$\mu$ problem in MSSM. The superpotential in the model
is~\cite{NMSSM}
\begin{equation}
W_{NMSSM}=\hat{Q}\hat{H}_{u}h_{u}\hat{U}^{C} +
\hat{H}_{d}\hat{Q}h_{d} \hat{D}^{C} +
\hat{H}_{d}\hat{L}h_{e}\hat{E}^{C} +
\lambda\hat{S}\hat{H}_{u}\hat{H}_{d} +\frac{1}{3}\kappa\hat{S}^{3},
\end{equation}
where $\kappa$ is a dimensionless constant and  measures the size
of Peccei-Quinn (PQ) symmetry breaking.

In addition to the two charged Higgs bosons, $H^{\pm}$, the physical NMSSM
Higgs sector consists of three scalars $h^0,\,H_{1,2}^0$ and two
pseudoscalars $A_{1,2}^0$.  As in the MSSM, $\tan\beta=v_u/v_d$ is
the ratio of the Higgs doublet vacuum expectation values $v_u=\langle
H_u^0\rangle=v\sin\beta$ and $v_d=\langle H_d^0\rangle=v\cos\beta$,
where $v=\sqrt{v_d^2+v_u^2}=\sqrt{2}m_W/g\simeq174GeV$. Generally,
the masses  and singlet contents of the physical fields depend
strongly on the parameters of the model (such as, in particular, how well the
PQ symmetry is broken). If the PQ symmetry is slightly broken, then
$A_1^0$ can be rather light, and its  mass  is given by
\begin{equation}
m^{2}_{A^{0}_{1}}=3\kappa x A_{k} +{\cal O}(\frac{1}{\tan\beta})
\end{equation}
with the vacuum expectation value of the
singlet $x=\langle S \rangle$; meanwhile, another pseudoscalar $A_{2}^0$ has a mass of
order of $m_{H^{\pm}}$.

For $\pi^{0}\to e^{+}e^{-}$ decay, the NMSSM contributions are
dominated by $A_{1}^0$. The couplings of $A_{1}^0$ to fermions
are~\cite{Hiller}
\begin{equation}
\mathcal {L}_{A_{i}^{0}f\bar{f}}=-i\frac{g}{2 m_{W}} \Big( X_{d}
m_{d} \bar{d}\gamma_{5}d + X_{u} m_{u} \bar{u}\gamma_{5}u + X_{\ell}
m_{\ell} \bar{\ell}\gamma_{5}\ell \Big) A_{1}^{0}
\label{Addcoupling1}
\end{equation}
where $X_{d}=X_{\ell}=\frac{v}{x}\delta_{-}$ and
$X_{u}=X_{d}/\tan^{2}\beta$; thus, the contribution of the
$\bar{u}\gamma_{5}u A_{1}^{0}$ term in $\pi^{0}\to e^{+}e^{-}$ could
be neglected in the large $\tan \beta$ approximation.

\begin{figure}[t]
\epsfxsize=7cm \centerline{\epsffile{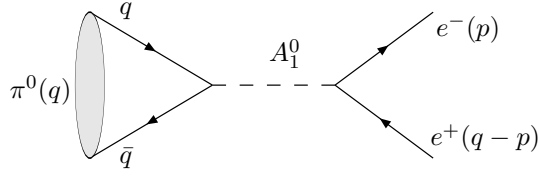}}
\centerline{\parbox{16cm}{\caption{\label{pitoeeNPfig} \small
Relevant Feynman diagram within NMSSM.}}}
\end{figure}

To the leading order, the relevant Feynman diagram within NMSSM is
shown in Fig.~\ref{pitoeeNPfig}. We obtain its amplitude as
\begin{equation}
\mathcal{M}_{A_{1}^{0}}=-\frac{G_{F}}{\sqrt{2}}m_{e}m_{\pi^0}^{3}f_{\pi^0}\frac{1}{m_{\pi^0}^{2}-m_{A_{1}^{0}}^{2}}X_d^{2},
\label{MA10}
\end{equation}
which  is independent of $m_{d}$, since
$m_{d}$ in the coupling of $A^{0}_{1}\bar{d}\gamma_{5}d$ is canceled by the
$m_d$ term of the hadronic matrix
\begin{equation}
\langle\,0\,|\,\bar{d}\gamma_{5}d\,|\,\pi^{0}\rangle\,=-\frac{i}{\sqrt{2}}f_{\pi^0} \frac{m_{\pi^{0}}^2}{2m_{d}}\,.
\end{equation}

\par
In the SM,  the normalized branching ratio of
$\pi^{0}\to e^{+}e^{-}$ is given by~\cite{Dorokhov}
\begin{equation}
R(\pi^{0}\to e^{+}e^{-})=\frac{{\cal B}(\pi^{0}\to
e^{+}e^{-})}{{\cal B}(\pi^{0}\to \gamma\gamma)} =2 \left(
\frac{\alpha_{e}}{\pi} \frac{m_{e}}{m_{\pi^0}} \right)^{2}
\beta_{e}(m^{2}_{\pi^0}) |{\cal A}(m^{2}_{\pi^0})|^{2}
\end{equation}
where $\beta_{e}(m^{2}_{\pi^0})=\sqrt{1-4\frac{m_e^2}{m_{\pi^0}^2}}$
and ${\cal A}(m^{2}_{\pi^0})$ is the reduced amplitude.

\par
To add the NMSSM amplitude to the above amplitudes consistently, we
rederive the SM amplitude to look into possible differences between
the conventions used in our Letter and the ones used in
Ref.~\cite{Dorokhov}.
\begin{figure}[ht]
\epsfxsize=6cm \centerline{\epsffile{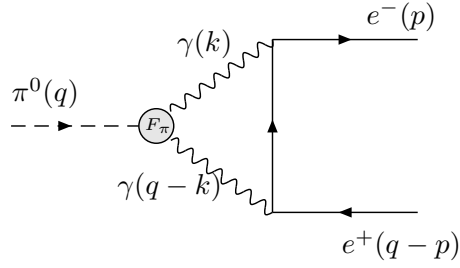}}
\centerline{\parbox{16cm}{\caption{\label{pitoeeSM} \small Triangle
diagram for $\pi^{0}\to e^+ e^-$ process.}}}
\end{figure}
 The Feynman diagram  that proceeds via two photon intermediate states is shown in Fig.~\ref{pitoeeSM}.
 We start with the $\pi^0\gamma^{*}\gamma^{*}$ vertex
\begin{equation}
H_{\mu\nu}=
-i\,e^{2}\,\epsilon_{\mu\nu\alpha\beta}\,k^{\alpha}\,(q-k)^{\beta}
f_{\gamma^{\ast}\gamma^{\ast}}\,F_{\pi^0\gamma^{\ast}\gamma^{\ast}}(k^2,(q-k)^2)\label{pieeVertex}
\end{equation}
where $k$ and $q-k$ are the momenta of the two photons,
$f_{\gamma^{\ast}\gamma^{\ast}}=\frac{\sqrt{2}}{4\pi^2\, and f_{\pi^0}}$
is the coupling constant of $\pi^{0}$ to two real photons.
$F_{\pi^0\gamma^{\ast}\gamma^{\ast}}(k^2,(q-k)^2)$ is the transition
form factor $\pi^0\to\gamma^{\ast}\gamma^{\ast}$, which is normalized to
$F_{\pi^0\gamma^{\ast}\gamma^{\ast}}(0,0)=1$. The  amplitude
of  Fig.~\ref{pitoeeSM} is written as
\begin{equation}
\mathcal {M}_{SM}(\pi^{0}\to e^+ e^-)=ie^2\int\frac{d^4k}{(2\pi)^4}
\frac{L^{\mu\nu}H_{\mu\nu}}{(k^2+i\varepsilon)\big((k-q)^2+i\varepsilon\big)\big((k-p)^2-m_{e}+i\varepsilon\big)},
\end{equation}
with
\begin{equation}
L^{\mu\nu}\,=\,\bar{u}(p,s)\gamma^{\mu}(\spur{p}-\spur{k}+m_e)\gamma^{\nu}v(q-p,s').
\end{equation}
There is a known, convenient way to calculate $L^{\mu\nu}$
with  the projection operator  for the outgoing
$e^+e^-$ pair system\cite{Martin}
\begin{eqnarray}
\mathcal{P}(q-p,p)&=&\frac{1}{\sqrt{2}}\big[v(q-p,+)\otimes\bar{u}(p,-)+v(q-p,-)\otimes\bar{u}(p,+)\big]\nonumber\\
                  &=&\frac{1}{2\sqrt{2t}}\big[-2m_e q_{\mu}\gamma^{\mu}\gamma^{5}+
\frac{1}{2}\epsilon_{\mu\nu\sigma\tau}\big{(}p^{\sigma}(q-p)^{\tau}-(q-p)^{\sigma}p^{\tau}\big{)}
\sigma^{\mu\nu}+t\gamma^{5}\big]
\label{Projectoree}
\end{eqnarray}
where $t=q^2=m_{\pi^0}^2$.  After some calculations, we get
\begin{equation}
\mathcal {M}_{SM}(\pi^{0}\to e^{+} e^{-})=
2\sqrt{2}\,\alpha^2\,m_e\,m_{\pi^0}\,f_{\gamma^{\ast}\gamma^{\ast}}\,A(m^{2}_{\pi})
\label{pieeSM}
\end{equation}
where the reduced amplitude $A(q^2)$ is
\begin{equation}
{\cal
A}(q^2)=\frac{2\,i}{q^2}\int\frac{d^4k}{\pi^2}\,\frac{k^2\,q^2\,-\,(q\cdot
k)^2}{(k^2+i\varepsilon)\big((k-q)^2+i\varepsilon\big)\big((k-p)^2-m_{e}+i\varepsilon\big)}
F_{\pi^0\gamma^{\ast}\gamma^{\ast}}(k^2,(q-k)^2). \label{pieeSMre}
\end{equation}
We note that the ${\cal A}(q^{2})$ derived here is in agreement with
Ref.~\cite{Dorokhov}. Further evaluation of the integrals of ${\cal
A}(q^{2})$ is quite subtle and lengthy~\cite{Bergstrom1,resconstr},
and only the imaginary part of ${\cal A}(m_{\pi^{0}}^{2})$ can be
obtained model-independently\cite{Drell,Bergstrom1}. In the
following calculations, we quote the result of Ref.~\cite{Dorokhov},
\begin{equation}
{\cal A}(m_{\pi}^{2})= (10.0\pm0.3) - i 17.5.
\end{equation}
With Eq.~(\ref{MA10}) and Eq.~(\ref{pieeSM}), we  get the total
amplitude
\begin{equation}
\mathcal{M}=
2\sqrt{2}\,\alpha^2\,m_e\,m_{\pi^0}\,f_{\gamma^{\ast}\gamma^{\ast}}\,A(m^{2}_{\pi})
-
\frac{G_{F}}{\sqrt{2}}m_{e}m_{\pi^0}^{3}f_{\pi^0}\frac{1}{m_{\pi^0}^{2}-m_{A_{1}^{0}}^{2}}X_d^{2}.
\end{equation}

\section{Numerical analysis and discussion}

\begin{figure}[ht]
\epsfxsize=8cm \centerline{\epsffile{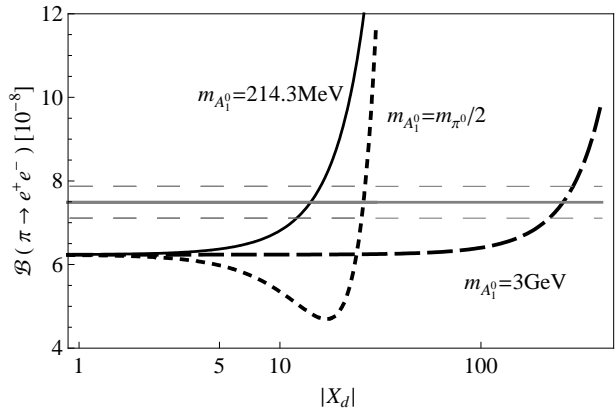}}
\centerline{\parbox{16cm}{\caption{\label{brld} \small The
dependence of ${\cal B}(\pi^{0}\,{\to}\,e^+ e^-)$ on the parameter
$|X_d|$ with $m_{A_1^0}=m_{\pi^0}/2$, $214.3 {\rm~MeV}$ and $\rm
3~GeV$, respectively. The horizontal lines are the KTeV data, where
the solid line is the central value and the dashed ones are the error
bars ($1\sigma$).}}}
\end{figure}

Now, we are ready to discuss the effects of $A_{1}^{0}$ numerically,
with a focus on the $m_{A^{0}_{1}}<2m_{b}$ scenarios. The
dependence of ${\cal B}(\pi^{0}\to e^+ e^-)$ on the parameter
$|X_d|$ is shown in Fig.~\ref{brld} with $m_{A_1^0}=m_{\pi}/2, \,
214.3{\rm MeV},\, 3{\rm GeV}$ as benchmarks. We have used the input
parameters ${\cal B}(\pi^{0}\to \gamma \gamma)=0.988$ and  $
f_{\pi^0}=(130.7\pm0.4)~{\rm MeV}$ ~\cite{PDG}.  As shown in Fig.~\ref{brld},
${\cal B}(\pi^{0}\to e^+ e^-)$ is very sensitive to the parameter
$|X_d|$ and $m_{A_1^0}$. For $m_{A_1^0}<m_{\pi^0}$, the NMSSM
contribution is deconstructive and reduces  ${\cal B}(\pi^{0}\to e^+
e^-)$ at small $|X_{d}|$ region. For $m_{A_1^0}>m_{\pi^0}$, the NMSSM
contribution is constructive and could enhance ${\cal B}(\pi^{0}\to
e^+ e^-)$ to be consistent with the KTeV measurement ${\cal
B}^{no-rad}_{KTeV}(\pi^{0}\to e^{+}e^{-})=(7.48\pm0.38)\times10^{-8}
$ (where $|X_{d}|$ strongly depends on $m_{A^{0}_{1}}$).
\\
{\bf I. Constraint on the  scenario of $m_{A^{0}_{1}}=214.3{\rm MeV}$}
\\
It is interesting to note that the HyperCP
Collaboration~\cite{HyperCP} has observed three events for the decay
$\Sigma^{+}\to p\mu^{+}\mu^{-}$ with a narrow range of dimuon
masses. This may indicate  that the decay proceeds via a neutral
intermediate state, $\Sigma^{+} \to p P^{0}, P^{0}\to \mu^{+}
\mu^{-}$, with a $P^{0}$ mass of $214.3\pm 0.5 {\rm MeV}$. The
possibility of $P^{0}$ has been explored in the
literature~\cite{He1, Valencia, He, Deshpande}.  The authors
 have proposed $A_{1}^{0}$ as a candidate for the $P^{0}$,
and have also shown that their explanation could be consistent with the constraints provided
by K and B meson decays~\cite{He1,He2}. It would be
worthwhile  to check on whether the explanation could be consistent with the
$\pi^{0}\to e^+ e^-$ decay.

Taking $m_{A_1^0}=214.3{\rm~MeV}$, we find that ${\cal B}(\pi^{0}\to
e^+ e^-)$ is enhanced rapidly and could  be consistent with the KTeV data within
$1\sigma$ for
\begin{equation}\label{xd214}
|X_d| =14.0\pm2.4.
\end{equation}
However,  the upper bound $|X_{d}|<1.2$ from the $a_{\mu}$
constraint has been derived  and used in the calculations of
Ref.~\cite{He1, He}. So, with the assumption that
$m_{A_1^0}=214.3{\rm~MeV}$, our result of $|X_{d}|$  violates the
upper bound with a significance of $5\sigma$.

Recently, CLEO~\cite{cleo} and BaBar~\cite{babar} have searched for
the CP-odd Higgs boson in radiative decays of $\Upsilon(1S)\to\gamma
A^{0}_{1}$  and $\Upsilon(3S)\to\gamma A^{0}_{1} $, respectively.
For $m_{A_1^0}=214{\rm MeV}$, CLEO gives the upper limit
\begin{equation}\label{xd2141S}
{\cal B}(\Upsilon(1S) \to \gamma
A_1^0)<2.3\times10^{-6}~~~~(90\%~C.L.)
\end{equation}
which constrains $|X_{d}| <0.16$.

The BaBar Collaboration has searched for $A_1^0$ through
$\Upsilon(3S) \to \gamma A_1^0$, $A_1^0\to \textrm{invisible}$ in
the mass range $m_{A_1^0}\leq7.8{\rm~GeV}$~\cite{babar}. From Fig.~5
of Ref.~\cite{babar},  we read
\begin{equation}\label{xd2141S2}
{\cal B}(\Upsilon(3S) \to \gamma A_1^0)\times{\cal B}(A_1^0\to
\textrm{invisible})\lesssim 3.5\times10^{-6}~~~~(90\%~ C.L.)
\end{equation}
for  $m_{A_1^0}=214{\rm~MeV}$. Assuming ${\cal B}(A_1^0\to
\textrm{invisible})\sim1$, we get the conservative upper limit
$|X_{d}| <0.19$

All of these upper limits are much lower than the limit of
Eq.\ref{xd214} set by $\pi^{0}\to e^{+}e^{-}$; therefore, the scenario where
$m_{A_1^0}\simeq214{\rm~MeV}$ in NMSSM could be excluded by
combining the constraints from $\pi^{0}\to e^{+}e^{-}$ and the
direct searches for $\Upsilon$ radiative decays.
\\
{\bf II. Constraints on the parameter space of $m_{A^{0}_{1}}-|X_{d}|$}

\begin{figure}[t]
\epsfxsize=16cm \centerline{\epsffile{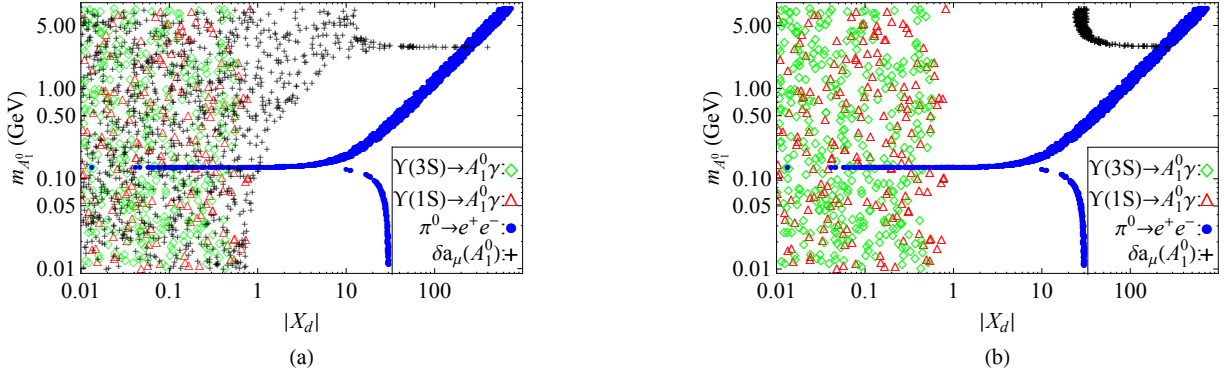}}
\centerline{\parbox{16cm}{\caption{\label{NPSpace1} \small
Constraints on the NMSSM parameter space through ${\cal
B}(\pi^{0}{\to}e^+ e^-)$, ${\cal B}(\Upsilon(1S) \to \gamma A_1^0)$,
${\cal B}(\Upsilon(3S) \to \gamma A_1^0)$ and $a_{\mu}$
respectively. The shaded regions are allowed by the labeled processes.}}}
\end{figure}

To show the constraints on NMSSM parameter space from $\pi^{0}\to
e^{+}e^{-}$, we present a scan of  $m_{A^{0}_{1}}-|X_{d}|$ space, as
shown in Fig.~\ref{NPSpace1}. In order to scan  the region of $m_{A_1^0}\sim
m_{\pi^0}$,  the amplitude of the $A_1^0$ contribution in
Eq.~(\ref{MA10}) is replaced by the Breit-Wigner formula
\begin{equation}\label{ammpi}
\mathcal{M}_{A_{1}^{0}}\,=-\,\frac{G_{F}}{\sqrt{2}}m_{e}m_{\pi^0}^{3}f_{\pi^0}\frac{1}{m_{\pi^0}^{2}
-m_{A_{1}^{0}}^{2}+i\Gamma(A_1^0)m_{A_1^0}}X_d^2\,.
\end{equation}
With the assumption that  $A_{1}^{0}$ just decays to electron and photon pairs
for $m_{A_1^0}\sim m_{\pi^0}$,  the decay width of $A_{1}^{0}$ could
be written as
\begin{equation}\label{gamma1}
\Gamma(A_1^0)=\Gamma(A_1^0\to e^+e^-)+\Gamma(A_1^0\to
 \gamma\gamma)
\end{equation}
with
\begin{eqnarray}\label{gamma2}
\Gamma(A_1^0\to e^+e^-)&=&\frac{\sqrt{2}G_{F}}{8\pi}m_{e}^2
m_{A_1^0}X_d^2\sqrt{1-4\frac{m_e^2}{m_{A_1^0}^2}}\,,\nonumber\\
 \Gamma(A_1^0\to\gamma\gamma)&=&\frac{G_{F}\alpha^2}{8\sqrt{2}\pi^3}m_{A_1^0}^{3}
 X_{d}^{2}|\sum_{i}rQ_{i}^2k_iF(k_i)|^2,
\end{eqnarray}
where $r=1$ for leptons and $r=N_c$ for quarks,
$k_i=m_i^2/m_{A_1^0}^2$ and $Q_i$ is the charge of the fermion in the loop. The
loop function $F(k_i)$ reads~\cite{Gunion}
\begin{equation}\label{F}
F(k_i)=\left\{
\begin{array}{ll}
-2\big(\arcsin\frac{1}{2\sqrt{k_i}}\big)^2\,~~~~~~~~~~~~~~\textrm{for}\,k_i\geq\frac{1}{4}\,,\nonumber\\
\frac{1}{2}\Big[\ln\big(\frac{1+\sqrt{1-4k_i}}{1-\sqrt{1-4k_i}}\big)+i\pi\Big]^2\,~~~~~~\textrm{for}\,k_i<\frac{1}{4}\,.
\end{array}
\right.
\end{equation}

As shown in Fig.~\ref{NPSpace1}, only two narrow connected bands of
the $|X_{d}|-m_{A^{0}_{1}}$ space survive after the KTeV measurement
of ${\cal B}(\pi^{0}{\to}e^+ e^-)$, which show  that $\pi^{0}{\to}e^+
e^-$ is very sensitive to  NP scenarios  with a light pseudoscalar
neutral boson.

In the following, we will determine which part of the remaining parameter space could satisfy the constraints enforced by radiative
$\Upsilon$ decays and $a_{\mu}$ simultaneously.

To include the $a_{\mu}$ constraint, we use the  experimental
result that~\cite{Bennett} $a_{\mu}(Exp)=(11659208.0\pm6.3)\times10^{-10}$
and the SM prediction~\cite{Passera}
$a_{\mu}(SM)=(11659177.8\pm6.1)\times10^{-10}$. The discrepancy is
\begin{equation}\label{amudisp}
\triangle\,a_{\mu}=a_{\mu}(Exp)-a_{\mu}(SM)=(30.2\pm8.8)\times10^{-10}(3.4\sigma)
\end{equation}
which is established at a $3.4\sigma$ level of significance.

The contributions of $A_1^0$ to $a_{\mu}$  are given by~\cite{Ellwanger}
\begin{eqnarray}\label{g2}
{\delta}a_{\mu}({A_1^0})&=&{\delta}a_{\mu}^{1-loop}({A_1^0})+{\delta}a_{\mu}^{2-loop}({A_1^0})\,,\\\nonumber
{\delta}a_{\mu}^{1-loop}({A_1^0})\,&=&\,-\sqrt{2}G_F
\frac{m_{\mu}^2}{8\pi^2}|X_d|^2f_{1}\big(\frac{m_{A_1^0}^2}{m_\mu^2}\big)\,,\\\nonumber
{\delta}a_{\mu}^{2-loop}({A_1^0})\,&=&\,\sqrt{2}G_F\alpha\frac{m_{\mu}^2}{8\pi^3}|X_d|^{2}
\Big[\frac{4}{3}\frac{1}{\tan^2\beta}f_{2}
\big(\frac{m_t^2}{m_{A_1^0}^2}\big)+\frac{1}{3}f_{2}
\big(\frac{m_b^2}{m_{A_1^0}^2}\big)+f_{2}\big(\frac{m_\tau^2}{m_{A_1^0}^2}\big)\Big]
\end{eqnarray}
with
\begin{eqnarray}\label{func}
 f_1(z)&=&\int_0^1dx\frac{x^3}{z(1-x)+x^2}\,,\\\nonumber
 f_2(z)&=&z\int_0^1dx\frac{1}{x(1-x)-z}\ln\frac{x(1-x)}{z}.
\end{eqnarray}
It has been found that the $A^{0}_{1}$ contribution is always negative at the one loop level
and worsens  the discrepancy in $a_{\mu}$; however, it could be
positive  and dominated by the two loop contribution for
$A^{0}_{1}>3{\rm GeV}$\cite{Ellwanger}. One should note that
there are other contributions to $a_{\mu}$ in NMSSM; for instance, the
chargino/sneutino and neutralino/smuon loops. Moreover, the discrepancy
$\triangle a_{\mu}$ could be resolved without
pseudoscalars~\cite{Ellwanger}.  So, putting a constraint on $|X_{d}|$ via  $a_{\mu}$
is a rather model-dependent process.
 There are two approximations with different emphases on the role of $A^{0}_{1}$;
namely, (i) assuming  that $\triangle a_{\mu}$ is resolved by other
contributions and requiring that $A_1^0$ contributions are smaller than the
$1\sigma$ error-bar of the experimental measurement, and (ii) assuming that the $A_1^0$
contributions are solely responsible for $\triangle a_{\mu}$.
In Ref.~\cite{He1}, approximation (i) has been used to derive an upper bound of
$|X_{d}|<1.2$. We present the $a_{\mu}$
constraints with the two approximations which are shown in Figs.~\ref{NPSpace1}(a)
and (b), respectively.

From Fig.~\ref{NPSpace1}(a), we can find that there are two narrow
overlaps between the constraints provided by $a_{\mu}$ and  ${\cal
B}(\pi^{0}{\to}e^+ e^-)$: one is for $m_{A_1^0}\sim 3{\rm~GeV}$ with
$|X_{d}|>150$ and another one is for $m_{A_1^0}\sim 135{\rm~MeV}$
with $|X_{d}|<1$.

In the searches for $\Upsilon\to\gamma A^{0}_{1}$ decays,
CLEO~\cite{cleo} obtains the upper limits for the product of ${\cal
B}(\Upsilon(1S)\to \gamma A^{0}_{1})$ and ${\cal
B}(A^{0}_{1}\to\tau^{+}\tau^{-})$ or ${\cal
B}(A^{0}_{1}\to\mu^{+}\mu^{-})$, while BaBar presents upper limits
on ${\cal B}(\Upsilon(3S)\to \gamma A^{0}_{1})\times {\cal
B}(A^{0}_{1}\to invisible)$. All these limits fluctuate with the
mass of $A^{0}_{1}$ frequently. For simplicity, we take  the loosest
upper limit  ${\cal B}(\Upsilon(1S)\to \gamma A^{0}_{1})\times {\cal
B}(A^{0}_{1}\to\tau^{+}\tau^{-})<6\times 10^{-5}$ of CLEO and assume
${\cal B}(A^{0}_{1}\to\tau^{+}\tau^{-})=1$. Similarly, we also use
the loosest upper limits  on ${\cal B}(\Upsilon(3S)\to\gamma
A^{0}_{1})\times {\cal B}(A^{0}_{1}\to invisible)<3.1\times 10^{-5}$
of BaBar~\cite{babar} and  assume ${\cal B}(A^{0}_{1}\to
invisible)=1$. With the loosest upper limits, we get their bounds on
the $|X_{d}|-m_{A^{0}_{1}}$ space, which are shown in
Fig.~\ref{NPSpace1}. From the figure, we can see the bounds
(excluding the parameter space $X_{d}>1$) for
$0<m_{A^{0}_{1}}<7.8{\rm~GeV}$. Fig.~\ref{NPSpace1}(b) shows that
there is no region of parameter space satisfying all the
aforementioned constraints if the contribution of $A^{0}_{1}$ is
required to solely resolve the $a_{\mu}$ discrepancy.

Of particular interest, as shown in Fig.~\ref{NPSpace1}(a), is the
parameter space around $m_{A_1^0}\sim 135{\rm~MeV}$ with $|X_{d}|<1$
(which is still allowed with  approximation (i)). To make a thorough investigation of the space,
 we read off the upper limits of BaBar~\cite{babar} from
Fig.~5  for the value $m_{A_1^0}\sim 135{\rm~MeV}$: ${\cal
B}(\Upsilon(3S) \to \gamma A_1^0)\times{\cal B}(A_1^0\to
invisible)\lesssim3.3\times10^{-6}$. With the assumption that  ${\cal
B}(A_1^0\to invisible)\simeq1$ and the constraints from ${\cal
B}(\pi^{0}{\to}e^+ e^-)$, we get
\begin{equation}\label{strairesul}
|X_d|=0.10\pm0.08,~~~~m_{A_1^0}=134.99\pm0.01\,{\rm MeV},
\end{equation}
where  the constraint on
$m_{A_1^0}$  is dominated by ${\cal B}(\pi^{0}{\to}e^+ e^-)$ and
 the  limit of $|X_d|$ is dominated by
  ${\cal B}(\Upsilon(3S) \to \gamma A_1^0)$. At first sight, the
uncertainties in the above–mentioned two parameters are too different. We
find  that the difference arises from our assumption $\Gamma(A_1^0)\simeq
\Gamma(A_1^0\to e^+e^-)+\Gamma(A_1^0\to\gamma\gamma)$. From
Eqs.~(\ref{ammpi}) and (\ref{gamma1}), one can see that the $X_{d}^{2}$
factor in ${\cal M}_{A_1^0}$ could be canceled out by the one in
$\Gamma(A_1^0)$  when $m_{A_1^0}$ approaches $m_{\pi^0}$, which
results in a very sharp peak for position of $m_{A_1^0}$. Thus, with
the well measured quantities given in Eq.~(\ref{ammpi}) and the sensitivity of
the peak, $m_{A_1^0}$  turns out to be well-constrained. Furthermore,
if we take $m_{A_1^0}= m_{\pi^0}$, we find that $X_{d}^{2}$ is canceled out
exactly, so there is no parameter to tune; however,  we  have
${\cal B}(\pi^{0}\to e^+ e^-)\gg1 $, which violates the unitary bound
and is thus excluded.

From the results of Eq.\ref{strairesul},  we obtain
${\delta}a_{\mu}({A_1^0})=(-9.2\pm8.9)\times 10^{-12}$ with
$\tan\beta=30$ as a benchmark,  which is small enough to be smeared
by the chargino/sneutrino and neutralino/smuon contributions.
Moreover, we have
\begin{equation}\label{gammaA10}
\Gamma(A_1^0)=(5.7\,\pm\,5.5)\times10^{-13}\,{\rm MeV},
\end{equation}
which corresponds to $\tau(A_1^0)\sim1.2\times10^{-9}~s$ ($c\tau \sim 36{\rm~cm}$).

\begin{figure}[t]
\epsfxsize=7cm \centerline{\epsffile{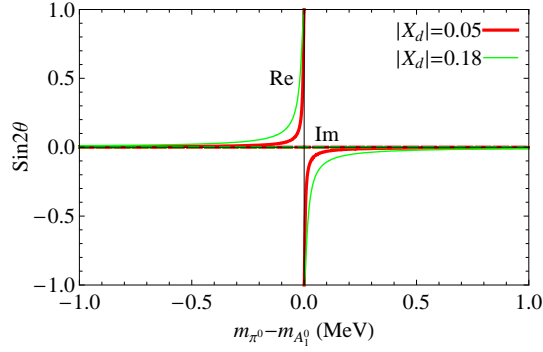}}
\centerline{\parbox{16cm}{\caption{\label{sin2theta}$\sin2\theta$
versus the mass difference of the unmixed states with $|X_d|=0.05$
and $0.18$. The solid and the dashed lines denote the real and the  imaginary
parts of $\sin2\theta$, respectively. \small}}}
\end{figure}

For the case where $A_1^0$ decays  mostly to invisible particles,
 we  take the width of $A_1^0$  as a free parameter and  get
$\Gamma(A_1^0)\leq8.24 \times\,10^{-6}{\rm GeV}$,
$m_{A_1^0}=134.99\pm0.02 {\rm~MeV}$ and $|X_d|\leq0.18$. In this
case, $m_{A_1^0}$ can equal $m_{\pi^{0}}$, and it is found that
$\Gamma(A_1^0)\leq3.3\times 10^{-6}{\rm~GeV}$ and $|X_d|\leq0.18$.
\\
{\bf III. The resonant effects of $m_{A^{0}_{1}}\sim m_{\pi^{0}}$}

So far we have included only the width effects of $A^{0}_{1}$ with the Breit-Wigner formula for the propagator of
$A^{0}_{1}$. When the  masses of $A_1^0$ and $\pi^0$ are very close,  the mixing between the two states
could modify the parton level  $\pi^{0}-A^{0}_{1}$ coupling.  In a manner analogous to Ref.\cite{mixing},
the mixing can be described by  introducing off-diagonal elements in the $A_1^0-\pi^0$ mass
matrix
\begin{eqnarray}
{\cal M}^2= \left(\begin{array}{cc}
m_{A_1^0}^2-i m_{A_1^0}\Gamma_{A_1^0} & \delta{m^2}\\
\delta{m^2} & m_{\pi^0}^2-im_{\pi^0}\Gamma_{\pi^0}
\end{array}\right)
\end{eqnarray}
with $\delta{m^2}=\sqrt{G_F/4\sqrt{2}}f_{\pi^0}m_{\pi^0}^2X_d$. The complex
mixing angle $\theta$ between the states is  given by
\begin{equation}
\sin^22\theta=\frac{(\delta
m^2)^2}{\frac{1}{4}(m_{A_1^0}^2-m_{\pi^0}^2-im_{A_1^0}\Gamma_{A_1^0}+im_{\pi^0}\Gamma_{\pi^0})^2+(\delta
m^2)^2}.
\end{equation}
The mass  eigenstates $A_1^{\prime0}$ and $\pi^{\prime0}$ are
obtained  as
\begin{eqnarray}
 A_1^{\prime0}&=&\frac{1}{N}(A_1^{0}\cos\theta+\pi^0\sin\theta),\\
 \pi^{\prime0}&=&\frac{1}{N}(-A_1^{0}\sin\theta+\pi^0\cos\theta),
\end{eqnarray}
where $N=\sqrt{|\sin\theta|^2+|\cos\theta|^2}$.  Then, we can  write
the decay amplitude of the ``physical'' state $\pi^{\prime0}$
 as
\begin{equation}
|{\cal M}(\pi^{\prime0} \to e^+
e^-)|^2\,=\,\frac{1}{N^2}\Big(|\cos\theta|^2|{\cal M}(\pi^{0} \to
e^+ e^-)|^2+|\sin\theta|^2|{\cal M}(A_1^{0} \to e^+ e^-)|^2\Big).
\end{equation}
Obviously, we obtain the SM result when $\theta$ is small.

With $|X_d|=0.05$ and $0.18$, Fig.~\ref{sin2theta} shows
$\sin2\theta$ as a function of the difference between $m_{A_1^0}$ and $m_{\pi^{0}}$.
We note that the imaginary part of $\sin2\theta$ is  negligibly  small,
since $\Gamma_{A^{0}_{1}} m_{A^{0}_{1}}+\Gamma_{\pi^{0}}m_{\pi^{0}}\ll\delta m^{2}$. So, the
normalization parameter  $N$ of the mixing states is nearly unity. Combining the
constraints from ${\cal B}(\Upsilon(3S) \to \gamma A_1^0)$ and
${\cal B}(\pi^{\prime0} \to e^+ e^-)$, we get
\begin{equation}
|X_d|=0.17\pm0.01,~~~~m_{A_1^0}\simeq m_{\pi^0}\,.
\end{equation}
This confirms the results of our straightforward calculation
from Eq.~(\ref{strairesul}), but gives a somewhat stronger constraint
on $|X_d|$. With this constraint,  we get
\begin{equation}
\Gamma(A_1^0)=(9.8\,\pm\,1.1)\times10^{-13}\,{\rm MeV},
\end{equation}
which is also in agreement with Eq.~(\ref{gammaA10}). Furthermore,
we get $|\sin\theta|^2=0.31\pm0.19$.

It is well known that the decay width of $\pi^{0}\to \gamma\gamma$ agrees perfectly with the SM prediction,
so it is doubtful that that $\pi^{0}\to \gamma\gamma$ would be compatible with Higgs with  a degenerate mass $m_{\pi^{0}}$.
 Using the fitted result $|\sin\theta|^2=0.31\pm0.19$ and
\begin{equation}
|{\cal M}(\pi^{\prime0} \to \gamma\gamma)|^2\,=\,\frac{1}{N^2}\Big(|\cos\theta|^2|{\cal M}(\pi^{0} \to
\gamma\gamma)|^2+|\sin\theta|^2|{\cal M}(A_1^{0} \to \gamma\gamma )|^2\Big),
\end{equation}
one can easily observe that
\begin{equation}
|{\cal M}(A_1^{0} \to \gamma\gamma )|^2 \simeq  |{\cal M}(\pi^{0} \to\gamma\gamma)|^2
\end{equation}
is needed to give $\Gamma(\pi^{\prime}\to \gamma\gamma)\simeq\Gamma(\pi^{0}\to \gamma\gamma)$.
However, it  would require a too large value of $|X_{d}|\simeq 10^{3}$; therefore, the degenerate case is excluded.

\section{Conclusion}
We have studied the decay $\pi^{0} \to e^+ e^-$ in the NMSSM and
shown that it is  sensitive to  the light CP-odd Higgs boson
$A^{0}_{1}$ predicted in the model. The possible  discrepancy
between the KTeV Collaboration measurement~\cite{KTeV} and the
theoretical prediction of  ${\cal B}(\pi^{0}\,{\to}\,e^+ e^-)$ could
be resolved in NMSSM by the effects of  $A^{0}_{1}$ at the tree
level. However, it excludes  a large fraction of the parameter space
of $m_{A^{0}_{1}}-|X_{d}|$. To further constrain  the parameter
space, we have included bounds from muon $g-2$ and the recent
searches for $A^{0}_{1}$ from radiative $\Upsilon$ decays performed
at CLEO~\cite{cleo} and BaBar~\cite{babar}. Combining all these
constraints, we have found that

\begin{itemize}
\item ${\cal B}(\pi^{0}\,{\to}\,e^+ e^-)$ and ${\cal B}(\Upsilon \to \gamma A_1^0)$ put strong constraints on
the NMSSM parameter $X_d$ and $m_{A_1^0}$. Due to their different
dependences on the two parameters,  the  interesting  scenario where
$m_{A^{0}_{1}}=214.3{\rm~MeV}$ is excluded, which would invalidate the
$A_1^0$ hypothesis for the three HyperCP events~\cite{HyperCP}.

\item Although these constraints point to  a pseudoscalar with $m_{A_1^0}\sim m_{\pi^{0}}$ and
$|X_d|=0.10\pm0.08$ ($0.17\pm0.01$,  $\pi^{0}-A_{1}^{0} $ mixing
included) in the NMSSM, such an $m_{A_1^0}$ is excluded by
$\pi^{0}\to\gamma\gamma$ decay.
\end{itemize}

In this Letter, we have worked in the limit of $X_{d}\gg X_{u}$, i.e., the large $\tan\beta$ limit.  If we relax
 the limit and take Eq.\ref{Addcoupling1}  as a general parameterization of the couplings
 between a pseudoscalar  and fermions, the $\bar u -u-A_{1}^{0}$ coupling should be included. However, its contribution is deconstructive to the contributions from $X_{d}$, since the $\pi^{0}$ flavor structure is
 $(u{\bar u} -d{\bar d})$. To give a result in agreement with the  KTeV Collaboration measurement~\cite{KTeV},
  $X_{u}\gg X_{d}$ would be needed, which would imply
possible large effects in $\Psi(1S)$ radiative decays. Detailed discussion of this issue would be
 beyond the main scope of our present  study. In summary, we could not find a region of parameter space of NMSSM with $m_{A_1^0}<7.8{\rm GeV}$ in the large $\tan\beta$ limit that is consistent with the experimental  constraints. The HyperCP $214.3{\rm MeV}$ resonance and the possible $3.3\sigma$ discrepancy in $\pi^{0}\to e^{+}e^{-}$ decay are still unsolved.
 Finally, further theoretical investigation is also needed to confirm the
discrepancy between the KTeV measurements and SM predications of $\pi^{0}\to e^{+}e^{-}$ decay. If the discrepancy still persists, it would be an important testing ground for NP scenarios with a light  pseudoscalar boson.

The work is supported by the National Science Foundation under contract
Nos.10675039 and 10735080.

 \end{document}